\documentclass[11pt,a4paper]{article}

\usepackage{jcappub}

\newcommand{\md}{\mathrm{d}}

\title{Structure Formation in the Effective Field Theory of Holographic Dark Energy}
\author{Alexander Ganz,}
\author{Chunshan Lin}
\affiliation[a]{ Faculty of Physics, Astronomy and Applied Computer Science, Jagiellonian University, 30-348 Krakow, Poland}

\emailAdd{alexander.ganz@uj.edu.pl}
\emailAdd{chunshan.lin@uj.edu.pl}

\abstract{ We investigate the structure formation in the effective field theory of the holographic dark energy. The equation of motion for the energy contrast $\delta_m$ of the cold dark matter is the same as the one in the general relativity up to the leading order in the small scale limit $k\gg aH$, provided the equation of state is Quintessence-like. Our effective field theory breaks down while the equation of state becomes phantom-like. We propose a solution to this problem by eliminating the scalar graviton. 
}

\arxivnumber{}

\notoc 

\begin{document}

\maketitle
\flushbottom

\section{Introduction}
The cosmological constant problem \cite{Weinberg:1988cp,Li:2011sd} is one of the most long-standing problem in physics, and has been attracting great attentions since the discovery of the cosmic accelerated expansion in 1998 \cite{SupernovaSearchTeam:1998fmf,SupernovaCosmologyProject:1998vns}. Why the cosmological constant is so small, while the quantum corrections seem to push the vacuum energy all the way to the $\rho_\Lambda\sim\Lambda^4$, where $\Lambda$ is the UV cut-off scale of the theory. The vacuum energy diverges quartically if we simply take the limit $\Lambda\to\infty$. It implies that we need to go beyond the Einstein's general relativity and/or the standard model in the particle physics to find a solution. 
One of the proposals, holographic dark energy (HDE) \cite{Cohen:1998zx,Li:2004rb}, is based on the holographic principle \cite{tHooft:1993dmi} relating the dark energy density to the boundary surface of the system with size $L$
\begin{align}
    \rho_\Lambda \sim M_p^2 L^{-2}
\end{align}
where $M_p^2= 8 \pi G$ is the reduced Planck mass. Requiring that the vacuum energy density of the system should be bounded by the mass of a black hole 
\begin{align}
    L^3 \Lambda^4 \leq L M_p^2,
\end{align}
we obtain the important UV-IR correspondence
\begin{align}
    \Lambda \sim \sqrt{M_P/L}.
\end{align}
The quartic divergence is cut off at the scale associated with the size of the physical system, which can be very small if we apply this idea to the universe as a whole. 
In \cite{Li:2004rb} it has been proposed that the IR cutoff is given by the future event horizon
\begin{align}
    R_H = a \int_t^\infty \frac{\md t^\prime}{a(t^\prime)}
\end{align}
which leads to an accelerated expansion of the Universe. While the model is in good agreement with current observations (see \cite{Wang:2016och} for a recent review) a covariant action has been missing. Recently, in \cite{Lin:2021bxv} this point has been addressed by providing a low-energy effective field theory (EFT) generalizing a first attempt in \cite{Li:2012xf} working only on the minisuperspace. The EFT is based on four St\"uckelberg fields leading to a Lorentz violating massive gravity theory with three degrees of freedom (two tensor modes and one scalar mode). 

The study on the influence of the holographic dark energy on the local physics has been missing in the literature. The EFT of the holographic dark energy has made it possible, well-motivated and very timely. While in \cite{Delgado:2021lzi} the impact on inflation is investigated, in this paper we want to study the phenomenological consequences of the EFT focusing on the impact of the holographic dark energy on the matter growth of dark matter by studying the linear perturbations around the Friedmann-Lemaitre-Robertson-Walker (FLRW) background. We will analyze both the quasistatic as the oscillating modes coming from the coupling of matter with HDE \cite{Bamba:2011ih}. We find that at the leading order in the momentum expansion, the equation of motion for the growth of the matter energy density contrast $\delta_m$ is the same as the one in the general relativity, while the corrections only appear at the higher order level. 

However, the linear perturbations of the EFT break down, i.e. become unstable, as soon as we enter the phantom regime $\omega_{de}<-1$ \cite{Lin:2021bxv}. It is common in phantom dark energy models due to the violation of the dominant dark energy condition \cite{Vikman:2004dc}. It may not be a pathology for our phenomenological studies, as long as the phantom transition did not occur neither in the past, nor at the present time (see \cite{Colgain:2021beg} for a recent discussion of the time of the turning point). While the transition may occur in the future, which would lead to the gradient instability -- an UV catastrophe. One might try to solve this problem by adding additional higher derivative operators similar to the approach in Horndeski models \cite{Deffayet:2010qz,Matsumoto:2017qil}. Instead, we will propose a different idea by eliminating the scalar degree of freedom by imposing additional constraints at the Hamiltonian level leading to a minimal modified gravity model with just two tensor modes \cite{Yao:2020tur}. By a careful choice of the constraints we can leave the background evolution untouched and only impact the scalar perturbations.
\medskip

The structure of the paper is as follows: In section \ref{sec:EFT_HDE} we will briefly recap the formulation of the EFT and discuss the validity range of the EFT.
From there we will derive the linear perturbations around the FLRW background of the HDE coupled with dust and discuss the stability properties in the small scale limit $k/(aH) \gg 1$ (section \ref{sec:Linear_Perturbations}). In section \ref{sec:Matter_growth} we will study impact of the HDE on dust first by using the quasistatic approximation and later by only relying on the small scale limit approximation. 
In the second part of the paper (section \ref{sec:Minimally_HDE}) we will discuss how we can eliminate the instability problem in the phantom regime by imposing additional constraints at the Hamiltonian and study two exemplary cases and the phenomenological consequences on the evolution of dust at the linear level. We will conclude the paper and present a short outlook in section \ref{sec:Conclusion}.

\section{EFT of Holographic Dark Energy}
\label{sec:EFT_HDE}
The covariant effective field theory of holographic dark energy can be expressed as \cite{Lin:2021bxv}
\begin{align}
\label{eq:action_holographic_dark_energy}
    S_{\mathrm{holo}} = \int \md^4x \, \sqrt{-g} \big[ \frac{M_p^2}{2} \mathcal{R} - M_p^2 \varphi^{-2} \big( (c+ \lambda) Z + \lambda \partial^\mu \varphi \partial_\mu \varphi + \frac{3 d}{8 Z} \bar \delta Z^{ab} \bar \delta Z^{cd} \delta_{ac} \delta_{bd} \big) \big],
\end{align}
where $\mathcal{R}$ is the 4-dimensional Ricci scalar, $\lambda$ a Lagrange multiplier, $\delta_{ab}$ the metric of the flat and SO(3) invariant field space and $Z= Z^{ab} \delta_{ab}$,
\begin{align}
    Z^{ab} =& \partial_\mu \phi^a \partial^\mu \phi^b - \frac{\partial_\mu \varphi \partial^\mu \phi^a \partial_\nu \varphi \partial^\nu \phi^b}{\partial_\alpha \varphi \partial^\alpha \varphi}, \\
    \bar \delta Z^{ab}=& Z^{ab} - 3 \frac{Z^{ac} Z^{bd} \delta_{cd}}{Z}.
\end{align}
The four scalar fields acquire non-trivial VEVs which break the spatial and temporal diffeomorphism invariance leading to the excitation of four Goldstone bosons
\begin{align}
    \phi^a = \frac{1}{\sqrt{3}} \left( x^i \delta_i^a +\pi^a \right), \qquad \varphi= \varphi(t) + \pi^0,
\end{align}
where $\delta_i^a$ is the pullback between the spacetime and fieldspace, $\pi^a$ are the spatial and $\pi^0$ the temporal Namubu Goldstone bosons.

\subsection{Background EOM}
At the FLRW background
\begin{align}
    \md s^2 = - \md t^2 + a^2(t) \md x^i \md x^j \delta_{ij}
\end{align}
the equations of motion (EOMs) can be written as
\begin{align}
    3 M_p^2 H^2 =& \frac{c}{a^2 \varphi^2} + \frac{\lambda }{2 a^4} + \rho_m, \\
    - M_p^2 \dot H =& \frac{c}{3 a^2 \varphi^2} + \frac{\lambda}{3 a^4} + \frac{1}{2} (\rho_m + p_m), \\
    \dot \varphi =& - \frac{1}{a} \qquad \dot \lambda = - \frac{4 c a}{\varphi^3},
\end{align}
 where $\rho_m$ and $p_m$ are the energy density and pressure of the matter part. Further, we have rescaled $\lambda \rightarrow \lambda \varphi^2/ (4 a^2 M_p^2)$ and $c\rightarrow c/M_p^2$. In the following we will consider pure dust for the matter part $p_m=0$ and set $M_p^2=1$ to simplify the notation.  
 
 Integrating the EOM of the St\"uckelberg field, $ \varphi $, leads to 
 \begin{align}
     \varphi(t) = \int_t^\infty \frac{1}{a(t^\prime)} \md t^\prime + \varphi(\infty),
 \end{align}
 where we have chosen to integrate up to the positive infinite future instead of the past. In \cite{Li:2012xf} it has been shown that $\varphi(\infty)=0$ and  consequently, $R_H= a \varphi$ is the size of the future event horizon.  
 
The equation of state of the dark energy, including the holographic component and the dark radiation, is given by
\begin{align}
\label{eq:Equation_of_state}
    \omega_{\mathrm{dark}} = \frac{p_{\mathrm{dark}}}{\rho_{\mathrm{dark}}} = \frac{\lambda \varphi^2 - 2 c a^2}{3 \lambda \varphi^2 + 6 c a^2}.
\end{align}
In \cite{Li:2012xf} the background EOMs has been solved analytically. Depending on the value of $c$ in the infinite future the dark energy behaves as a cosmological constant $c=6$, crosses into the phantom regime $c< 6$ or behaves as a Quintessence model $c>6$.  In \cite{li2013cosmological} the authors constrained the free parameters by using the data from Union2.1+ BAO+ CMB+ $H_0$ showing that a phantom equation of state is highly favored by the current cosmological data. 

\begin{figure}
    \includegraphics[scale=0.55]{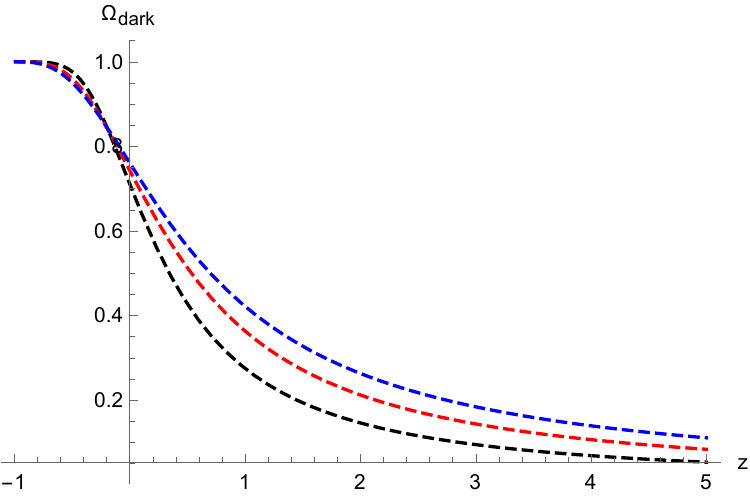} \hfill 
    \includegraphics[scale=0.55]{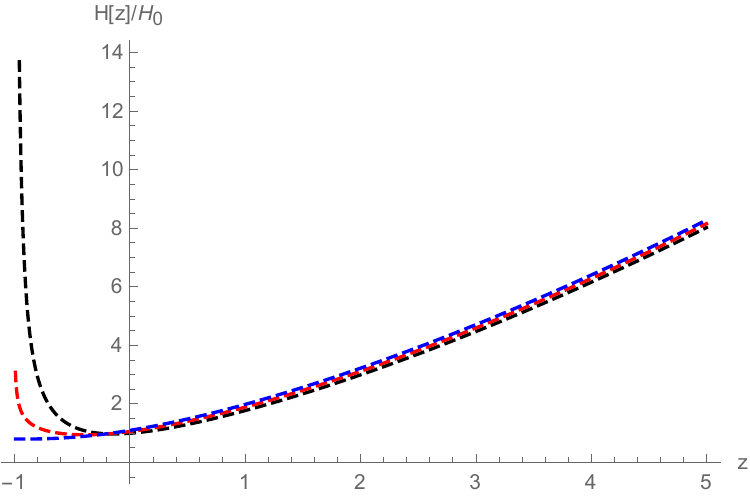} \\
    \includegraphics[scale=0.55]{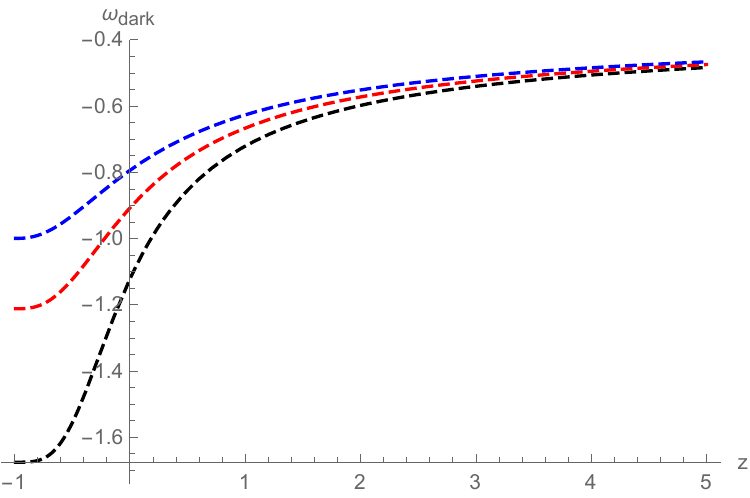} \hfill 
    \includegraphics[scale=0.55]{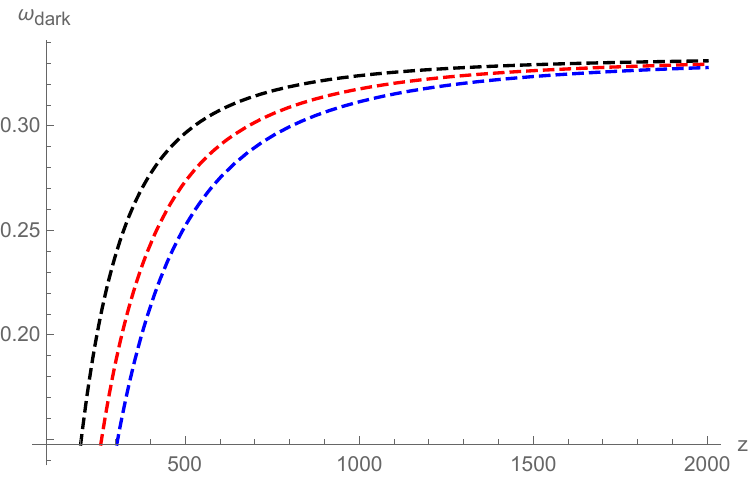}
    \caption{Numerical solution of $\Omega_{\mathrm{dark}}$ , $H(z)/H_0$, and $\omega_{\mathrm{dark}}$  for a set of three different initial conditions which are given by $c=2.23,\, \tilde \varphi=0.7$ (black color) $c=4,\,\tilde\varphi=0.938$ (red color) and $c=6,\,\tilde\varphi=1.148$ (blue color).}
    \label{fig:Background_Evolution}
\end{figure}

To get an impression of the different cosmological scenarios we plot in figure \ref{fig:Background_Evolution}  the evolution of the normalized dark energy energy density, $\Omega_{\mathrm{dark}} = \rho_{\mathrm{dark}}/3 H^2$, the Hubble parameter and the equation of state at the late and early times for a different set of initial values. As one scenario we choose the best fit parameters from the aforementioned reference. For the other scenarios we adapt the values of $c$ and $\tilde \varphi=H_0 \varphi$, where $H_0$ is the current Hubble parameter, but keeping the ratio of $c/\tilde\varphi^2$ constant in order to shift the time of entering into the phantom regime or avoid it all. 
As discussed, we can see that for $c<6$ the equation of state crosses into the phantom regime leading to a big rip in the future, while for $c=6$ the dark energy component behaves a cosmological constant. Furthermore, in all of the cases, at early times the dark energy component behaves as dark radiation, $\omega_{\mathrm{dark}}\simeq 1/3$.

\subsection{Validity of the EFT}
The UV-IR correspondence in the holographic dark energy implies that, if we take the future event horizon as our IR cutoff, the UV cutoff varies as the universe expands. For the HDE to be relevant to the cosmological study, it is required that
\begin{align}
    \Lambda > H.
\end{align}
Defining the energy density of the holographic part as
\begin{align}
    \Omega_{\mathrm{hde}} = \frac{\rho_{\mathrm{hde}}}{3 H^2} =\frac{1}{3 H^2} \frac{c}{a^2 \varphi^2} \sim \frac{1}{H^2 R_H^2},
\end{align}
and using $\Lambda \sim \sqrt{M_p/R_H}$ we can note that the previous condition translates to 
\begin{align}
  \Omega_{\mathrm{hde}} > \frac{H^2}{M_p^2}.
\end{align}
We set $t_0$ as the time of nowaday and $a(t_0) = 1$. Further, we have $\Omega_{\mathrm{hde}}(t_0) \sim \mathcal{O}(1) $ and $H_0^2(t_0)/M_p^2 \sim 10^{-120}$. Tracing it backward to the matter radiation equality moment
\begin{align}
    \Omega_{\mathrm{hde}}(t_{eq}) \sim &\, \frac{\rho_{\mathrm{hde}}(t_0) a^{-2}}{\rho_c(t_0) a^{-3}} \sim 10^{-3}, \\
    \frac{H^2(t_{eq})}{M_p^2} \sim &\, 10^9 \cdot 10^{-120},
\end{align}
the inequality is still valid up to the radiation matter equality moment. Tracing it further backwards to the radiation domination epoch
\begin{align}
    \Omega_{\mathrm{hde}} \sim &\, \Omega_{\mathrm{hde}}(t_{eq}) \left( \frac{a}{a_{eq}} \right)^2 \sim 10^{-3} \cdot \left( \frac{a}{a_{eq}} \right)^2, \\
    \frac{H^2}{M_p^2} \sim &\, 10^{-111} \left( \frac{a_{eq}}{a} \right)^4,
\end{align}
and comparing both relations we obtain that the inequality $\Omega_{\mathrm{hde}} > H^2/M_p^2$ breaks down when
\begin{align}
    a \sim 10^{-18} a_{eq} \sim 10^{-21} \sim e^{-48}.
\end{align}
Therefore, the EFT is valid for the most part of the hot big bang history, namely about 48 e-folding numbers counting backward from the present time.

\section{Linear Perturbations}
\label{sec:Linear_Perturbations}

In the following we will consider the linear perturbations around the FLRW background. We will use the gauge freedom to fix the perturbations of the St\"uckelberg fields to zero, $\pi^0,\pi^a=0$. For the scalar perturbations of the metric we choose the convention
\begin{align}
    g_{00} =& 1 + 2\alpha + \alpha^2, \\
    g_{0i} =& a \partial_i \beta, \\
    g_{ij} =& a^2 \left[ (1 + 2\psi ) \delta_{ij} + \partial_i \partial_j E \right].
\end{align}

\subsection{Dust}
For the matter part we choose the Schulz-Sorkin action \cite{schutz1977variational} following the implementation in \cite{Felice_2016}
\begin{align}
    S_{\mathrm{mat}} =& - \int \md^4x\, \left( \sqrt{-g} \rho_m(n) + J^\alpha \partial_\alpha \varphi \right), \\
    \rho_m(n) =& \mu_0 n, \\
    n=& \sqrt{\frac{J^\alpha J^\beta g_{\alpha\beta}}{g}},
\end{align}
where $J^\alpha$ is a vector of weight one and  $\varphi$, $n$ and $\rho_m$ are scalar fields. The perturbations of the dust can be expressed as
\begin{align}
    J^0 =& \mathcal{N}_0 + \delta j_0, \\
    J^k = & \delta^{kj} \partial_j \delta j, \\
    \varphi = & - \mu \int^t \md \tau N(\tau) - \mu_0 v_m,
\end{align}
where $\mathcal{N}_0$ is an integration constant, namely the number of dust particles and at the background level $a^3 \rho_m = \mu_0 \mathcal{N}_0  $.

In Fourier space we obtain for the second order of the dust action 
\begin{align}
    \delta S_{\mathrm{mat}}= \int \md t \md^3 k \mu_0 \Big[ \frac{1}{2} \mathcal{N}_0 k^2 \beta^2 + \frac{k^2}{a^2} \delta j v_m + \frac{k^2}{a} \beta \delta j+ \frac{1}{2 \mathcal{N}_0} \frac{k^2}{a^2} \delta j^2 + \dot v_m \delta j_0 - \alpha \delta j_0 \Big].
\end{align}
As a next step, it is useful to replace $\delta j_0$ with the gauge invariant matter overdensity $\delta_m$ via
\begin{align}
     \frac{\delta j_0}{\mathcal{N}_0} = \delta_m - 3 H v_m + 3 \psi - \frac{1}{2} k^2 E.
\end{align}
Further, we can solve the EOM for $\delta j$ which leads to
\begin{align}
    \delta j = - (v_m + a \beta ) \mathcal{N}_0.
\end{align}
Finally, the second order dust action can be expressed as
\begin{align}
    \delta S_{\mathrm{mat}}= \int \md t \md^3 k\, a^3 \rho_m & \Big[ - \frac{1}{2} \frac{k^2}{a^2} v_m^2 - \frac{1}{2} k^2 E \dot v_m + \frac{3}{2} \dot H v_m^2  + \frac{1}{2} k^2 E \alpha  + 3 H v_m \alpha \nonumber \\
    &- \frac{k^2}{a } v_m \beta + \delta_m \dot v_m - \alpha \delta_m + 3\psi \dot v_m - 3 \alpha \psi \Big].
    \label{eq:Second_order_action_dust}
\end{align}

\subsection{Second order action}
Using the definitions of the metric perturbations and the background EOMs the second order action for the holographic dark energy \eqref{eq:action_holographic_dark_energy} is given by
\begin{align}
    \delta S_{\mathrm{holo}} = \int \md^3k \md t a^3 &  \Big[  2 H \alpha \frac{k^2 \beta}{a} - \frac{d k^4 E^2 }{36 a^2 \varphi^2 } - \delta \lambda \frac{k^2 E + 6 \alpha - 6 \psi }{12 a^4} - 2 \dot \psi  \frac{k^2 \beta}{a} \nonumber \\
    & + \frac{k^2}{a^2} \psi (2 \alpha + \psi ) - \frac{\lambda }{48 a^4} \left( k^4 E^2 + 4 k^2 E (\alpha + \psi ) -12 ( \alpha + \psi)^2 \right) \nonumber \\
    & - \frac{c}{12 a^2 \varphi^2} \left( k^4 E^2 + 4 k^2 E (\alpha + \psi) - 12 \psi (2\alpha +\psi) \right)   \nonumber \\
    & - \frac{1}{2} \rho_m \alpha (k^2 E - 6 \psi) - ( H \alpha - \dot \psi) \left( 3 H \alpha + k^2 \dot E - 3 \dot \psi \right) \Big],
\end{align}
where we again rescaled $\lambda\rightarrow \lambda \varphi^2/(4 a^2)$ and similar for $\delta \lambda$.
Combining it with \eqref{eq:Second_order_action_dust} we can integrate out the  scalar fields $\delta \lambda$, $\delta N$, $\beta$, $E$ and $v_m$.
Further, it is convenient to redefine the metric variables to diagonalize the kinetic metric and to rescale the matter perturbations via 
\begin{align}
   \delta_m =& k \tilde \delta_m, \\
    \psi =& \tilde \psi + \frac{k a^4 \rho_m \varphi^2}{2 (\lambda \varphi^2 + a^2 (c + k^2 \varphi^2 ) )} \tilde \delta_m.
\end{align}
The action can then be expressed as
\begin{align}
    \delta S = \int \md^3k \md t\, a^3 \Big[ \beta_1 \dot{\tilde\psi}^2 + \beta_2 \dot{\tilde \delta}_m^2 +  \beta_3 \dot{\tilde \psi} \tilde\delta_m  + \beta_4 \psi^2 + \beta_5 \delta_m^2 + \beta_6 \tilde\psi \tilde\delta_m \Big],
    \label{eq:Perturbed_action}
\end{align}
where in the small scale limit, $ x\equiv k/a H \gg 1$,
\begin{align}
    & \beta_1 = -\frac{c+d}{a^2 H^2 \varphi^2} + \mathcal{O}(x^{-2}),  \\
    &\beta_2 = \frac{1}{2} \rho_m a^2 + \mathcal{O}(x^{-2}), \\
    &\beta_3 = \frac{2(2c+d) \rho_m }{k H \varphi^2 }+ \mathcal{O}(x^{-3}), \\
    &  \beta_4 = -  \frac{k^2}{a^2} \left(\frac{\lambda }{3 a^4 H^2} + \frac{c}{3 H^2 \varphi^2 a^2} \right)+ \mathcal{O}(x^0),  \\
    &  \beta_5=  \frac{1}{4} a^2 \rho_m^2 + \mathcal{O}(x^{-2}),  \\
    &\beta_6 = \rho_m \frac{2 c a^2 H + (2c+d) \rho_m a^3 \varphi - 2 (4c+d) a^3 H^2 \varphi}{k a^3 H^2 \varphi^3} + \mathcal{O}(x^{-3}),
\end{align}
wwhere $\mathcal{O}(x^n)$ is standing for the next-to-leading order terms which are not relevant for our later purposes. 

Let us make some comments about the structure of the linear perturbations at leading order in the small scale limit. The stability conditions for the absence of ghost and gradient instabilities for $\tilde \psi$ are the same as in \cite{Lin:2021bxv} without the presence of dust, namely
\begin{align}
    b\equiv& - 2 ( c+d) >0, \\
    c_s^2\equiv& -   \frac{\lambda \varphi^2 + c a^2 }{3  (c+d) a^2}=\frac{c}{b}\cdot\frac{\rho_{\text{dark}}}{\rho_{\text{hde}}}\cdot\left(1+w_{\text{dark}}\right) >0.
\end{align}
We have used the equation of state of the dark component eq. \eqref{eq:Equation_of_state} to get the above relation. We can note that $c_s^2 \ge 0$ $\leftrightarrow$ $\omega_{\mathrm{dark}} \ge -1$. Therefore, the perturbations become unstable as soon as we enter the phantom regime $\omega_{\mathrm{dark}} < -1$.

The absence of ghost instabilities in the matter sector requires $\rho_m >0$ which is fulfilled for any standard canonical matter, while the sound speed is, as expected, vanishing. In general, we can note that the two degrees of freedom decouple from each other at leading order in the small scale limit since the interaction terms are all suppressed.

\section{Matter growth}
\label{sec:Matter_growth}
From the action \eqref{eq:Perturbed_action} the corresponding EOMs are given by
\begin{align}
    - \frac{1}{a^3}\frac{\md }{\md t} \left( a^3  \beta_1 \dot{\tilde\psi} \right) - \frac{1}{2}  \beta_3 \dot{\tilde \delta}_m +   \beta_4 \tilde\psi + \frac{1}{2} \big[ \beta_6 - \frac{1}{a^3} \partial_t ( a^3 \tilde \beta_3 )  \big] \tilde\delta_m = 0, \label{eq:EOM1} \\
    - \frac{1}{a^3}\frac{\md }{\md t} \left( a^3  \beta_2 \dot{\tilde\delta}_m \right) + \frac{1}{2}  \beta_3 \dot{\tilde \psi}_m +  \beta_5 \tilde\delta_m + \frac{1}{2} \beta_6 \tilde\psi = 0. \label{eq:EOM2}
\end{align}
To evaluate the impact of the holographic dark energy on the growth of the matter structure we will employ the quasistatic approximation. In the appendix \ref{app:Oscillating_modes} we explicitly check that the oscillating modes coming from the coupling of the holographic dark energy to the dust modes are indeed subdominant.  

Using the quasistatic approximation, namely
$\ddot{\tilde\psi} \sim H \dot{\tilde\psi} \sim H^2 \tilde\psi \ll (k^2/a^2) \tilde \psi$, the first EOM simplifies to
\begin{align}
    \beta_4 \tilde\psi  = \frac{1}{2} \beta_3 \dot{\tilde \delta}_m - \frac{1}{2} \big[ \beta_6 - \frac{1}{a^3} \partial_t ( a^3 \tilde \beta_3 )  \big] \tilde\delta_m. \label{eq:Poisson_equation} 
\end{align}
Note, that it leads to $\tilde \psi \propto \mathcal{O}(x^{-3}) \tilde \delta_m$.
Plugging it back into the second EOM and only consider the leading order in the small scale expansion we obtain
\begin{align}
    \ddot{\tilde\delta}_m + 2 H \dot{\tilde\delta}_m- \frac{1}{2} \rho_m \tilde\delta_m =0.
\end{align}
Therefore, up to leading order  in $\mathcal{O}(x)$, the EOM for the matter growth is the same as the  standard result from GR.
\medskip

Similarly, by using 
\begin{align}
    \Psi =& \alpha + \frac{\md ( a\beta)}{\md t} - \frac{\md (a^2 \dot E) }{\md t} , \label{eq:Psi_Bardeen} \\
    \Phi =& - \psi - H a \beta + a^2 H \dot E,
    \label{eq:Phi_Bardeen}
\end{align}
we can derive the Poisson equation for the two gauge invariant Bardeen potentials. Applying the quasistatic approximation we obtain, as expected, up to leading order
\begin{align}
    - \frac{k^2}{a^2} \Psi = \frac{1}{2} \rho_m \delta , \\
    - \frac{k^2}{a^2} \Phi = \frac{1}{2} \rho_m \delta.
\end{align}
We recover the standard Poisson equation and the absence of gravitational slip, $\eta = \Psi/ \Phi =1$. 

Therefore, the presence of the holographic dark energy only impacts the growth of the matter structure indirectly due to a change of the background evolution of the Hubble parameter similar to Quintessence models. A detailed analysis is beyond the scope of this paper and we postpone it to a future project.

\section{Minimally holographic dark energy models}
\label{sec:Minimally_HDE}
In the previous section we have seen that the effective field theory of holographic dark energy breaks down as soon as we enter the phantom regime. This problem is not unique for this model but instead a common problem of linear perturbations in scalar field theories with a phantom regime due to the violation of the dominant energy condition \cite{Vikman:2004dc}. One possible approach to solve this issue might be to enlarge the model by introducing additional higher derivative operators which could solve the instability problem similar to the discussion in Horndeski models \cite{Deffayet:2010qz,Matsumoto:2017qil}.  

In this paper we want to shortly present an alternative approach by eliminating the scalar degree of freedom by introducing additional constraints via Lagrange multiplier. In \cite{Yao:2020tur} it has been discussed how the scalar degree of freedom can be removed by imposing further constraints at the Hamiltonian level. By choosing constraints which vanish identically on the FLRW background we can keep the full background dynamics but instead only remove the scalar degree of freedom at linear and higher order. In the following we will present two different choices of possible constraints. We will limit our discussion to the phenomenological consequences and refer to the aforementioned reference for the theoretical details.

Using the ADM decomposition of the metric
\begin{align}
    \md s^2 = - N^2 \md t^2 + h_{ij} ( \md x^i + N^i \md t) (\md x^j + N^j \md t),
\end{align}
we can express the Hamiltonian of the holographic dark energy model \eqref{eq:action_holographic_dark_energy}
 as \cite{Lin:2021bxv}
 \begin{align}
     H_T = \int \md^3x\, \Big[ \mathcal{H} + N^k \mathcal{H}_k + u_N \pi_N + u_i \pi^i + u_a p^a + u_\lambda p_\lambda \Big],
 \end{align}
 where  
 \begin{align}
     \mathcal{H} =& N \mathcal{H}_{GR} + N \sqrt{h} \frac{c + \lambda}{\varphi^2} Z + \sqrt{h} \frac{\lambda}{\varphi^2} \frac{\dot \varphi^2}{N} + \sqrt{h} N \frac{3 d}{8 \varphi^2 Z} \bar \delta Z^{ab} \bar \delta Z^{cd} \delta_{ac} \delta_{bd}, \\
     \mathcal{H}_k =& - 2 D_j \pi^j_k + p^a \partial_k \phi^b \delta_{ab}  + p_\lambda \partial_k \lambda,
 \end{align}
 with $\mathcal{H}_{GR}$ being the standard Hamiltonian constraint of GR, $\pi^{ij}$, $\pi_N$, $\pi^k$, $p^a$ and $p_\lambda$ are the momenta associated to $h_{ij}$, $N$, $N^k$, $\phi^a$ and $\lambda$ and $D_j$ is the covariant derivative with respect to the induced spatial metric $h_{ij}$. Further, we have used the gauge freedom to fix the temporal Goldstone boson $\varphi=\varphi(t)$ to be just a function of time so that
 \begin{align}
     Z^{ab} = D_k \phi^a D^k \phi^b.
 \end{align}
Note, however, that we have not fixed the spatial Goldstone boson to keep the spatial diffeomorphism invariance. Therefore, $\pi^k$ and $\mathcal{H}_k$ are the usual six first class constraints, while $\pi_N$, $p_\lambda$ and $p_a$, the momenta associated to $N$, $\lambda$ and $\phi_a$ are second class constraints which leads to 5 additional secondary constraints. Combined we have 6 first class and 10 second class constraints resulting in three degrees of freedom.

\subsection{Non-propagating model}
The easiest way to eliminate the scalar degree of freedom is by adding a constraint which does not depend on the momentum of the metric,
\begin{align}
    \tilde H_T = H_T + \int \md^3x\, u_R C_R,
\end{align}
with
\begin{align}
    C_R=\sqrt{h} f(h_{ij},D_k,t).
\end{align}
For simplicity, we choose $C_R = \sqrt{h} R$, where $R$ is the 3-dimensional Ricci scalar. However, as long as the constraint does not depend on the momentum and vanishes identically on the FLRW background the specific choice of the constraint does not impact the scalar part up to the linear order in perturbation theory. The conservation of the primary constraint $C_R$ enforces a secondary constraint 
\begin{align}
      C_{R,2} = - 4 N \tilde R_{ij} \tilde \pi^{ij}+ 4 \pi^{ij} D_i D_j N  \approx 0,
\end{align}
where $\tilde R_{ij} = R_{ij} - R\, h_{ij}/3$ is the traceless component and similar for $\tilde \pi^{ij}$.
Note, that both of these constraints are identities on the flat FLRW background and, therefore, the dynamics at the background remain unchanged \footnote{Note, that if we consider constraints like $C_R = \sqrt{h} R^2$ the constraint would be trivially fulfilled up to the linear level for a flat spatial background. Therefore, due to the missing constraint the scalar degree of freedom from the holographic dark energy would still be present at the linear level leading to an inconsistent perturbation theory.}. This is similar to the approach in \cite{DeFelice:2015hla,DeFelice:2015moy,Aoki:2020ila} where the imposed constraints at the FLRW background either vanish or are equivalent to the Bianchi identity. 

However, in our case, it also implies that the background value of the Lagrange parameter, $u_R$, is completely unconstrained by the EOMs. 
To make the last point more clear let us consider the Lagrangian. 
Since the primary constraint $C_R$ does not depend on the momentum the Legendre transformation becomes trivial and the action is given by
\begin{align}
    S = S_{\mathrm{holo}} + \int \md^3 x \sqrt{h} u_R R,
\end{align}
where $S_{\mathrm{holo}}$ is the standard action from the holographic part without the constraints (see eq. \eqref{eq:action_holographic_dark_energy}) and we have rescaled the Lagrange parameter $u_R \rightarrow - u_R$.
From the trace of the metric EOM we obtain 
\begin{align}
    D_k D^k u_R = \frac{h_{ij}}{2 \sqrt{h}} \frac{\delta S_{\mathrm{holo}} }{\delta h_{ij}}.
\end{align}
The Lagrange parameter can be obtained by solving the elliptic differential equation. Since the FLRW background does not depend on space the Lagrange parameter $u_R = u_{0}(t)$ will be completely unconstrained at the background level. On the other hand, at the full non-linear level $u_{0}(t)$ is specified by the boundary conditions of the spatial differential equation. 

\begin{figure}
    \centering
    \includegraphics[scale=0.55]{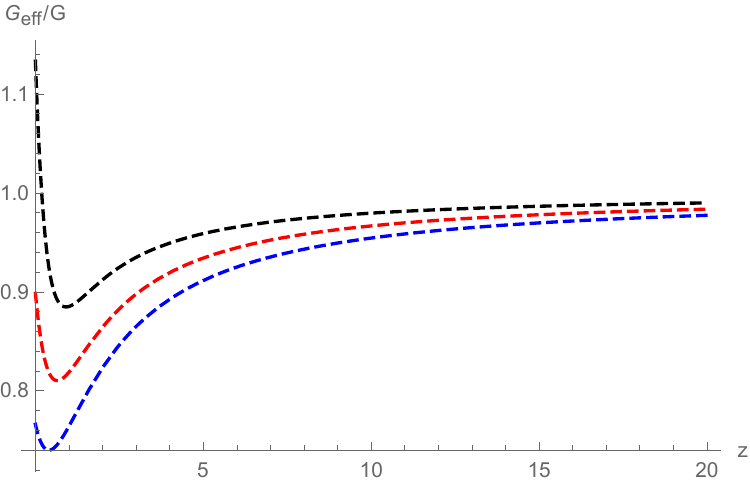}
    \caption{The evolution of the effective gravitational constant with the same initial conditions as in fig. \ref{fig:Background_Evolution}.}
    \label{fig:effective_Gravitational_constant}
\end{figure}

As a next step, let us consider the impact of the constraint at the linear level. Up to linear order the new constraint will lead to the condition
\begin{align}
    \delta R = -4 \partial^2 \psi =0.
\end{align}
Enforcing proper boundary conditions we set $\psi=0$. Therefore, following the same steps as in the previous section the scalar perturbations in the presence of dust can be written as
\begin{align}
    \delta S= \int \md^3k \md t\,\frac{a^3 \rho_m \varphi^2 \big[ a^3 \rho_m \delta_m + 2 a^3 H \dot \delta_m \big]^2}{4 a ((c+d) a^3 \rho_m + 2 a H^2 \lambda \varphi^2 + 2 a^3 H^2 (c+k^2 \varphi^2))},
\end{align}
where we have again rescaled $\lambda \rightarrow \lambda \varphi^2/ (4 a^2)$. Due to the presence of the constraints at linear order we have just one dynamical scalar degree of freedom. In the small scale limit $x=k/(aH) \gg 1$ the action simplifies to 
\begin{align}
    \delta S = \int \md^3 k\md t\, a^3 \rho_m \frac{a^2}{2k^2} \Big[ \dot \delta_m^2 - \left( \frac{\rho_m \lambda}{6 a^4 H^2} + \frac{c \rho_m }{6 a^2 \varphi^2 H^2} - \frac{\rho_m}{2}  \right) \delta_m^2 \Big].
\end{align}
Therefore, in the small scale limit there is no ghost instability and the dust does not propagate. 
The EOM can be written as
\begin{align}
    \ddot \delta_m + 2 H \dot \delta_m - \frac{1}{2} \rho_m \left( 1 - \frac{3}{2} \Omega_{\mathrm{dark}} (1 + \omega_{\mathrm{dark}})  \right) \delta_m =0.
\end{align}
where
\begin{align}
    \Omega_{\mathrm{dark}} = \frac{\rho_{\mathrm{dark}}}{3H^2} = \frac{1}{3 H^2} \left( \frac{c}{a^2 \varphi^2} + \frac{\lambda}{2 a^4} \right)
\end{align}
Consequently, the effective growth,
\begin{align}
    \frac{G_{\mathrm{eff}}}{G} =  1 - \frac{3}{2} \Omega_{\mathrm{dark}} (1 + \omega_{\mathrm{dark}}) ,
\end{align}
is suppressed as long as $\omega_{\mathrm{dark}} > -1$ and enhanced otherwise. In figure \ref{fig:effective_Gravitational_constant} we have plotted the effective gravitational constant. We can see that it could be used to put severe constraints on the model parameters which we postpone to a future project.
Using eqs. \eqref{eq:Phi_Bardeen} and \eqref{eq:Psi_Bardeen} the Poisson equation for the two Bardeen gravitational potentials is given by
\begin{align}
    - \frac{k^2}{a^2} \Psi = & \frac{1}{2} \frac{G_{\mathrm{eff}}}{G} \rho_m \delta_m, \\
    - \frac{k^2}{a^2} \Phi = & \frac{1}{2} \rho_m \delta_m. 
\end{align}
The gravitational slip, $\eta$, is then given by
\begin{align}
    \eta = \frac{\Psi}{\Phi} = \frac{G_{\mathrm{eff}}}{G} =1 - \frac{3}{2} \Omega_{\mathrm{dark}} (1 + \omega_{\mathrm{dark}} ).
\end{align}


\subsection{Propagating model}

An alternative way, is to directly impose two constraints which do not commute with each other. As before, the constraints should be trivially fulfilled for the background but not for higher order. 
For simplicity, let us consider the two constraints 
\begin{align}
    C_1=& D_k D^k \pi, \\
    C_2 =& R_{ij} \pi^{ij}.
\end{align}
The second constraint can be split into a product of the two traces and the traceless components. Since the traceless part $\tilde \pi^{ij}$ vanishes at the background  at the linear level $C_2$ will result in the same constraint as $C_R$, i.e. $\delta R=0$.

Using the Hamiltonian equations of motion we obtain
\begin{align}
    \dot h_{ij} = \{ h_{ij}, H_t\}\approx  \frac{2 N}{ \sqrt{h}} \left( 2 \pi_{ij} - \pi h_{ij} \right) + 2 D_{(i} N_{j)} + D_k D^k u_1 h_{ij} + u_2 R_{ij},
\end{align}
we can express the momentum of the metric in terms of the extrinsic curvature
\begin{align}
    \pi_{ij} = \frac{\sqrt{h}}{2}  \left( K_{ij} - K h_{ij}\right) +  \frac{\sqrt{h}}{2 N} h_{ij} D_k D^k u_1 +  \frac{\sqrt{h}}{4N} u_2 ( R h_{ij} - R_{ij} ).
\end{align}
Performing the Legendre transformation the action is finally given by
\begin{align}
    S = S_{\mathrm{holo}} +& \int \md^4x\,  \sqrt{h} \big[ \frac{1}{2} u_2 (K R- K_{ij} R^{ij}) + \frac{1}{8N} u_2^2 (R_{ij} R^{ij}-R^2) \nonumber \\
    &+ K D_k D^k u_1- \frac{1}{2 N}  u_2 R  D_k D^k u_1 - \frac{3}{4 N} D_k D^k u_1 D^jD_j u_1 \big].
\end{align}
 Let us make some comments. First, by construction the background value of $u_1$ and $u_2$ does not impact the background equations of motion. 
At linear level, the first constraint result in $k^2 \psi=0$ and the second one fixes $\delta u_1$
\begin{align}
   \frac{k^2}{a^2} \delta u_1 =   2 H \alpha - \frac{2}{3}\frac{k^2 \beta}{a} + \frac{1}{3} k^2 \dot E. 
\end{align}
Plugging it back into the action we obtain
\begin{align}
    \delta S_{\mathrm{new}} = \int \md^4x\,a^3 \frac{1}{3} (- \frac{k^2 \beta}{a} + 3 H \alpha + \frac{1}{2} k^2 \dot E)^2.
\end{align}
Combining it with the results from the holographic action we obtain in the subhorizon limit, $x \gg 1$,
\begin{align}
    \delta S = \int \md^3 k \md t\, a^3 \frac{a^2 \rho_m}{2 k^2 } \Big[  \dot \delta_m^2 - \left(  c_s^2 \frac{k^2}{a^2} + m^2  \right) \delta_m^2 \Big],
\end{align}
where
\begin{align}
    c_s^2 =& \rho_m \frac{ R_H^2}{b }, \\
    m^2 =& \rho_m \frac{-\frac{3}{2} b \rho_m \varphi^2 a^4 + 3 b H^2 \varphi^2 a^4  - b \lambda \varphi^2 - b c a^6 + 3 H a^2 \varphi ( - 2 b a^3  + 3 \rho_m H \varphi^3)  }{b^2 a^6 }.
\end{align}
Requiring the absence of ghost and gradient instabilities requires again $b>0$ and $\rho_m>0$. Due to the constraints we can change the fundamental structure of dust as providing a non-vanishing sound speed which puts severe constraints on $b$ in order to be consistent with the observation of dust-like dark matter. 

\begin{figure}
    \centering
    \includegraphics[scale=0.55]{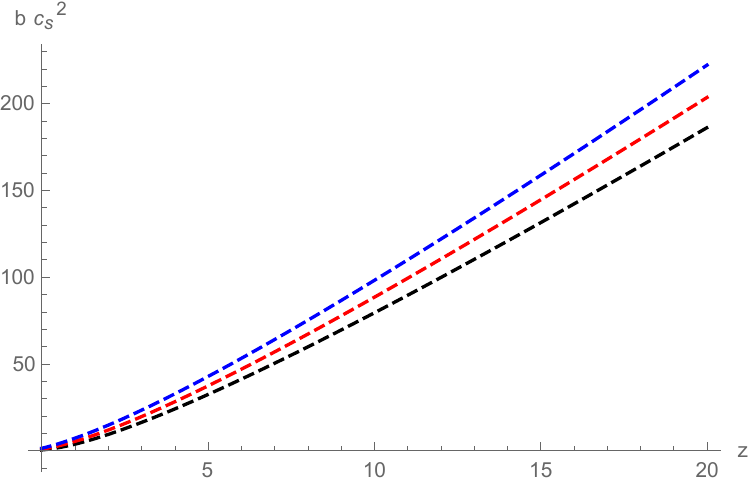}
    \caption{The evolution of the sound speed of the propagating dust with the same initial conditions as in fig. \ref{fig:Background_Evolution}.}
    \label{fig:sound_speed_constraint}
\end{figure}

We have plotted $b c_s^2$ in fig. \ref{fig:sound_speed_constraint}. We can see that around $z\simeq 15$ in all the cases the sound speed is of the order $c_s^2 \sim 10^2/b$. Since the sound speed of dark matter is highly constrained, $c_s^2 < 10^{-6} - 10^{-8}$ (assuming a modified Chaplyn gas) \cite{Avelino:2015dwa} it leads to severe lower bounds $b > 10^{8} - 10^{10}$, which may lead to an instability in the tensor sector. 

\section{Conclusion}
\label{sec:Conclusion}
In the paper we have discussed the impact of HDE on the structure of matter growth. Using the EFT derived in \cite{Lin:2021bxv}, we showed by using the quasistatic approximation that up to the leading order in the small scale limit $k/(aH) \gg 1$ we recover the standard behavior of GR. HDE only impacts the matter growth indirectly due to the changed background evolution of the Hubble parameter similar to Quintessence models. However, the description of the linear perturbations break down as soon as we are entering into the phantom regime, $\omega_{\mathrm{dark}} < -1$, leading to unstable linear perturbations. 

Therefore, in a second step, we discussed minimal HDE models. Using the concept proposed in \cite{DeFelice:2015hla} we eliminate the scalar degree of freedom at linear order by imposing additional constraints at the Hamiltonian. By choosing constraints which vanish on the FLRW background the constraints do not impact the background evolution. To analyze the phenomenological consequences we focused on two different examples of possible constraints. While eliminating the scalar degree of freedom at the linear order the minimal HDE still impacts the evolution of dust leading to a suppressed or enhanced matter growth, $\omega_{\mathrm{dark}} > -1$ or $\omega_{\mathrm{dark}} < -1$, or even change the fundamental structure of dust by providing a non-vanishing sound speed. Due to the strict constraints on the sound speed of the dark matter the latter case can be ruled out. The first case, however, can provide interesting models in order to address the Hubble and $\sigma_8$ tension at the same time. We leave this issue for future work.  

\acknowledgments
C.L. is supported by the grant No. UMO-2018/30/Q/ST9/00795 and A.G. by the grant No.   UMO-2021/40/C/ST9/00015 both from  the  National  Science  Centre, Poland.   We would like to thank Yin-zhe Ma, Sunny Vagnozzi, and Pierre Zhang for the useful discussions.

\appendix

\section{Oscillating modes}
\label{app:Oscillating_modes}
Let us analyze the EOMs without using the quasistatic approximation following \cite{Bamba:2011ih,Matsumoto:2018dim} to confirm that the oscillating modes are indeed subdominant.
From the two EOMs \eqref{eq:EOM1} and \eqref{eq:EOM2} we can derive a single fourth order differential EOM
\begin{align}
    \frac{\md^4 \tilde\delta_m}{\md t^4}  +  A_3 \frac{\md^3 \tilde\delta_m}{\md t^3}  +  A_2\frac{\md^2 \tilde\delta_m}{\md t^2} + A_1 \frac{\md \tilde\delta_m}{\md t} + A_0 \tilde\delta_m = 0,
\end{align}
where
\begin{align}
    A_3 =& \frac{2 c a + 9 c a^2 H \varphi + 11 H \lambda \varphi^3 }{c a^2 \varphi + \lambda \varphi^3} + \mathcal{O}(x^{-2}), \\
    A_2 = & c_s \frac{k^2}{a^2} + \mathcal{O}(x^0), \\
    A_1 =&  2 H c_s^2 \frac{k^2}{a^2} + \mathcal{O}(x^0), \\
    A_0 =& - \frac{1}{2} \rho_m c_s^2 \frac{k^2}{a^2} + \mathcal{O}(x^0).
\end{align}
Note, that the single fourth order differential equations of motion encompass both degrees of freedom. By considering the small-scale limit we can solve the four modes analytically. 
The first two solutions can be obtained by considering the highest order of the differential equation in $x$, which leads to
\begin{align}
   c_s^2 \frac{k^2}{a^2}\left( \frac{\md^2 \tilde \delta_m}{\md t^2} + 2 H \frac{\md \tilde \delta_m}{\md t} - \frac{1}{2} \rho_m \tilde\delta_m \right)=0.
\end{align}
These two solutions are equivalent to the ones which we have derived by using the quasistatic approximation. Note, that our expansion only works as long as $c_s k/aH \gg 1$. Therefore, it breaks down as soon as the sound speed approaches zero.

The other two solutions of the matter overdensity are oscillating. These are the modes which were neglected by using the quasistatic approximation and are coming due to the coupling of matter with holographic dark energy.  By using the WKB approximation
\begin{align}
    \tilde\delta_m = C(t) \exp( S(t) ),
\end{align}
we obtain 
\begin{align}
    S(t) = \pm i \int \md t \frac{c_s k}{a} +  S_0,
\end{align}
where $S_0$ is an integration constant and
\begin{align}
    \frac{\md}{\md t} \log C(t) = - \frac{1}{2} A_3 + \frac{A_1}{2 A_2} - \frac{5 \dot A_2}{4 A_2} = -  H - \frac{c a + H \lambda \varphi^3}{c a^2 \varphi + \lambda \varphi^3} - \frac{5}{2} \frac{\dot c_s}{c_s}.
\end{align}
Finally, restricting to real solutions the oscillating modes can be expressed as
\begin{align}
    \tilde \delta_m =  \frac{ \exp\left( - \int \md t\frac{c a + H \lambda \varphi^3}{c a^2 \varphi + \lambda \varphi^3} + S_0 \right)}{a c_s^{5/2}}\left[ c_1 \cos\left( \int \md t\frac{c_s k}{a} \right) + c_2 \sin\left( \int \md t\frac{c_s k}{a}  \right)  \right].
\end{align}
In order to check if these new oscillating modes are relevant we can evaluate the effective growth of these modes in comparison to the quasistatic modes following the discussion in \cite{Matsumoto:2018dim}. 

The effective growth rate can be defined as
\begin{align}
    f_{\mathrm{eff}}= \frac{\md \log C}{ \md \log a} = -1 - \frac{1}{H} \frac{c a + H \lambda \varphi^3}{c a^2 \varphi + \lambda \varphi^3} - \frac{5}{2} \frac{\dot c_s}{H c_s}.
\end{align}
Note, that we obtain a singularity problem if $c_s \rightarrow 0$. Therefore, our description only holds as long as $\omega_{\mathrm{dark}}> -1$.

\begin{figure}
    \includegraphics[scale=0.55]{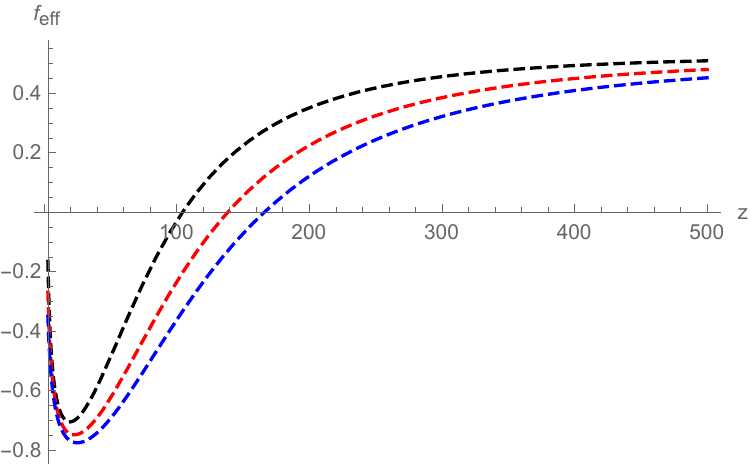} \hfill 
    \includegraphics[scale=0.55]{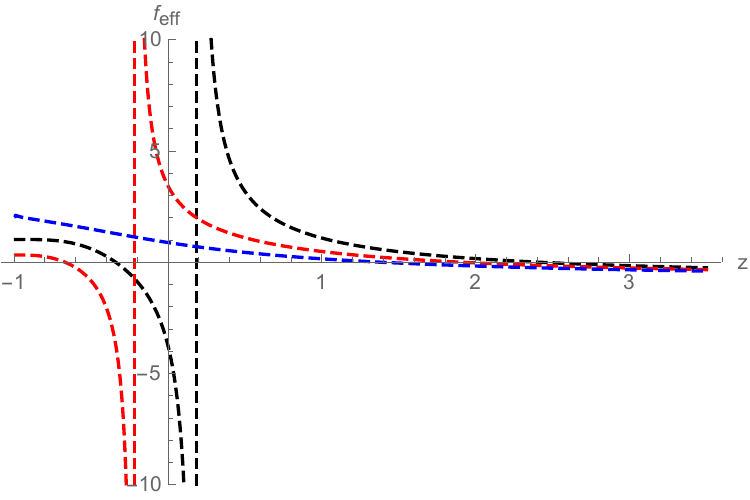}
    \caption{Evolution of the effective growth rate, $f_{\mathrm{eff}}$, with the same initial conditions as in fig. \ref{fig:Background_Evolution}.}
    \label{fig:effective_growth_rate}
\end{figure}

In figure \ref{fig:effective_growth_rate} we have plotted the evolution of the effective growth. On the left hand side the evolution for the early time from $z=3$ up to $z=500$ is shown. For $z<100$ in all of cases $f_{\mathrm{eff}} <0$. Therefore, the oscillating modes decay away. Since the quasistatic modes from standard GR grow in this time period with $f\simeq 1$ the oscillating modes can safely be neglected today, if we assume that the initial conditions for the different modes are of the same order of magnitude. However, there is a caveat as we can see on the right hand side of fig. \ref{fig:effective_growth_rate}, where we have plotted the effective growth rate from $z=3.5$ up to the far future $z=-1$. The effective growth rate of the oscillating solutions diverge when $\omega_{\mathrm{dark}}\rightarrow -1$ since $c_s^2\rightarrow 0$. However, we have to note that in this regime also the validity of our assumptions break down so that we cannot trust anymore the results of the effective growth rate near the crossing into the phantom regime. Therefore, we can conclude that in the regime, where we can trust our expansion, the oscillating solutions are subdominant in comparison to the quasistatic modes confirming the quasistatic approximation. 

\bibliography{bibliography}
\bibliographystyle{JHEP}

\end{document}